\documentclass[11pt, onecolumn, romanappendices, letterpaper]{IEEEtran}

\usepackage{xspace}
\usepackage{mathrsfs}
\usepackage{amsopn}

\def\var{\mathop{\rm Var}\nolimits}%
\newcommand{\Bc}{\mathcal{B}}
\newcommand{\Cc}{\mathcal{C}}

\newcommand{\Ec}{\mathcal{E}}

\newcommand{\Sc}{\mathcal{S}}

\newcommand{\Vc}{\mathcal{V}}

\newcommand{\Xc}{\mathcal{X}}
\newcommand{\Yc}{\mathcal{Y}}
\newcommand{\Zc}{\mathcal{Z}}

\newcommand{\Yv}{{\bf Y}}
\newcommand{\Zv}{{\bf Z}}

\newcommand{\Sv}{{\bf S}}

\newcommand{\sv}{{\bf s}}

\newcommand{\aep}{{\mathcal{T}_{\epsilon}^{(n)}}}

\newcommand{\Mh}{{\hat{M}}}

\newcommand{\lh}{{\hat{l}}}
\newcommand{\mh}{{\hat{m}}}

\newcommand{\Rt}{{\tilde{R}}}

\newcommand{\lt}{{\tilde{l}}}

\def\d{\delta}
\def\e{\epsilon}

\DeclareMathOperator\E{E}
\let\P\relax
\DeclareMathOperator\P{P}

\def\textiid{i.i.d.\@\xspace}
\newcommand\iid{\ifmmode\text{ i.i.d. } \else \textiid \fi}

\usepackage{cite}
\usepackage{version}
\usepackage{fullpage,epsfig,graphicx,psfrag,color,subfig}
\usepackage{amsmath}
\usepackage{amssymb}
\newtheorem{lemma}{Lemma}
\newtheorem{theorem}{Theorem}
\newtheorem{proposition}{Proposition}

\begin{document}
\title{Wiretap Channel with Causal State Information}

\author{
\authorblockN{Yeow-Khiang Chia and Abbas El Gamal}\\
\authorblockA{Department of Electrical Engineering\\
Stanford University\\
Stanford, CA 94305, USA\\
Email:  ykchia@stanford.edu, abbas@ee.stanford.edu}
}

\maketitle

\begin{abstract}
A lower bound on the secrecy capacity of the wiretap channel with state information available causally at both the encoder and decoder is established. The lower bound is shown to be strictly larger than that for the noncausal case by Liu and Chen. Achievability is proved using block Markov coding, Shannon strategy, and key generation from common state information. The state sequence available at the end of each block is used to generate a key, which is used to enhance the transmission rate of the confidential message in the following block. An upper bound on the secrecy capacity when the state is available noncausally at the encoder and decoder is established and is shown to coincide with the lower bound for several classes of  wiretap channels with state. 

\end{abstract}
\section{Introduction}
\begin{figure}[!ht]
\begin{center}
\small
\psfrag{A}[c]{$M$}
\psfrag{B}[c]{$$}
\psfrag{C}[c]{$X_i$}
\psfrag{D}[l]{$S_i$}
\psfrag{E}[c]{}
\psfrag{F}[c]{$Y_{i}$}
\psfrag{F2}[c]{$Z_{i}$}
\psfrag{G}[c]{$\Mh$}
\psfrag{G2}[c]{}
\psfrag{P1}[c]{\rm Encoder}
\psfrag{P4}[c]{\rm Decoder}
\psfrag{P5}[c]{\rm {\small Eavesdropper}}
\psfrag{P2}[c]{$p(s)$}
\psfrag{P3}[c]{$p(y,z|x,s)$}
\psfrag{s}[b]{}
\includegraphics[width=0.8\linewidth]{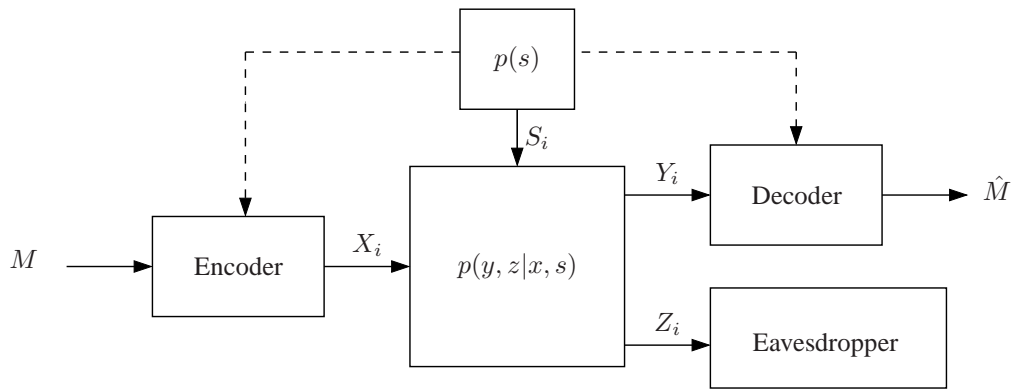} 
\caption{Wiretap channel with State} \label{fig1}
\end{center}
\end{figure}

Consider the 2-receiver wiretap channel with state depicted in Figure~\ref{fig1}. The sender $X$ wishes to send a message to the legitimate receiver $Y$ while keeping it asymptotically secret from the eavesdropper $Z$. The secrecy capacity for this channel can be defined under various scenarios of state information availability at the encoder and decoder. When the state information is not available at either party, the problem reduces to the classical wiretap channel for the channel averaged over the state and the secrecy capacity is known~\cite{Wyner},~\cite{Csiszar}. When the state is available only at the decoder, the problem reduces to the wiretap channel with augmented receiver $(Y,S)$. 

The interesting scenarios to consider therefore are when the state information is available at the encoder and may or may not be available at the decoder. This raises the question of how the encoder and decoder can make use of the state information to increase the secrecy rate. In ~\cite{Chen--Vinck2006}, Chen and Vinck established a lower bound on the secrecy capacity when the state information is available noncausally only at the encoder. The lower bound is established using a combination of Gelfand--Pinsker coding and Wyner wiretap coding. Subsequently, Liu and Chen~\cite{Liu--Chen2007} used the same techniques to establish a lower bound on the secrecy capacity when the state information is available noncausally at both the encoder and decoder. In a related direction, Khisti, Diggavi, and Wornell~\cite{Khisti--Diggavi--Wornell2009} considered the problem of secret key agreement first studied in~\cite{Maurer1993} and~\cite{Ahlswede--Csiszar1993} for the wiretap channel with state and established the secret key capacity when the state is available causally or noncausally at the encoder and decoder. The key is generated in two parts; the first using a wiretap channel code while treating the state sequence as a time-sharing sequence, and the second part is generated from the state itself. 

In this paper, we consider the wiretap channel with state information available {\em causally} at the encoder and decoder. We show that the lower bound for the noncausal case in~\cite{Liu--Chen2007} is achievable when only causal state information is available. Our achievability scheme, however, is quite different from that for the noncausal case. We use block Markov coding, Shannon strategy for channels with state~\cite{Shannon1958a}, and secret key agreement from state information, which builds on the work in~\cite{Khisti--Diggavi--Wornell2009}. However, unlike~\cite{Khisti--Diggavi--Wornell2009}, we are not directly interested in the size of the secret key, but rather in using the secret key generated from the state sequence in one transmission block to increase the secrecy rate in the following block. This block Markov scheme causes additional information leakage through the correlation between the secret key generated in a block and the received sequences at the eavesdropper in subsequent blocks. Although a similar block Markov coding scheme was used in~\cite{Ardestanizadeh--Franceschetti--Javidi--Kim2008} to establish the secrecy capacity of the degraded wiretap channel with rate limited secure feedback, in their setup no information about the key is leaked to the eavesdropper because the feedback link is assumed to be secure. 

We also establish an upper bound on the secrecy capacity of the wiretap channel with state information available noncausally at the encoder and decoder. We show that the upper bound coincides with the aforementioned lower bound for several classes of channels. Thus, the secrecy capacity for these classes does not depend on whether the state information is known causally or noncausally at the encoder.

The rest of the paper is organized as follows. In Section~\ref{sect:2}, we provide the needed definitions. In Section~\ref{sect:3}, we summarize and discuss the main results in the paper. The proofs of the lower and upper bounds are detailed in Sections~\ref{sect:4} and~\ref{sect:5}, respectively.
 
\section{Problem Definition} \label{sect:2}
Consider a discrete memoryless wiretap channel (DM-WTC) with discrete memoryless state (DM)\\ $(\Xc\times\Sc, p(y,z|x,s)p(s), \Yc, \Zc)$ consisting of a finite input alphabet $\Xc$, finite output alphabets $\Yc$, $\Zc$, a finite {\em state} alphabet $\Sc$, a collection of conditional pmfs $p(y,z|x,s)$ on $\Yc\times \Zc$, and a pmf $p(s)$ on the state alphabet $\Sc$. The sender $X$ wishes to send a confidential message $M \in [1:2^{nR}]$ to the receiver $Y$ while keeping it secret from the eavesdropper $Z$ with either causal or noncausal state information available at both the encoder and decoder.

A $(2^{nR}, n)$ code for the DM-WTC with causal state information at the encoder and decoder consists of: (i) a message set $[1:2^{nR}]$, (ii) an encoder that generates a symbol $X_i(m)$  according to a conditional pmf $p(x_i|m, s^{i}, x^{i-1})$ for $i\in[1:n]$; and a decoder  that assigns an estimate $\hat{M}$ or an error message to each received sequence pair $(y^n,s^n)$. We assume throughout that the message $M$ is uniformly distributed over the message set. The probability of error is defined as $P_{e}^{(n)} = \P\{\Mh \neq M\}$.
The information leakage rate at the eavesdropper $Z$, which measures the amount of information about $M$ that leaks out to the eavesdropper, is defined as $R_L = \frac{1}{n} I(M;Z^n)$. A secrecy rate $R$ is said to be achievable if there exists a sequence of codes with $P_{e}^{(n)} \to 0$ and $R_L \to 0$ as $n\to \infty$. The secrecy capacity $C_{\rm S-CSI}$ is the supremum of the set of achievable rates.

We also consider the case when the state information is available noncausally at the encoder. The only change in the above definitions is that the encoder now generates a codeword $X^n(m)$ according to the conditional pmf $p(x^n|m, s^{n})$, i.e., the stochastic mapping is allowed to depend on the entire state sequence instead of just the past and present state sequence. The secrecy capacity for this scenario is denoted by $C_{\rm S-NCSI}$.

The notation used in this paper will follow that of El Gamal--Kim Lectures on Network Information Theory~\cite{El-Gamal--Kim2010}.

\section{Summary of Main Results}\label{sect:3}

We summarize the results in this paper. Proofs are given in the following two sections and in the Appendix.

\subsection*{Lower Bound}
The main result in this paper is the following lower bound on the secrecy capacity of the DM-WTC with causal state information available causally at both the encoder and decoder.

\begin{theorem} \label{thm:1}
The secrecy capacity of the DM-WTC with state information available causally at the encoder and decoder is lower bounded as
\begin{align}
C_{\rm S-CSI} & \ge \max \{ \max_{p(v|s)p(x|v,s)} \min \{I(V;Y|S) - I(V;Z|S)+ H(S|Z), I(V;Y|S)\}, \nonumber \\ 
& \left. \qquad \qquad \max_{p(v)p(x|v,s)} \min \{H(S|Z,V), I(V;Y|S)\}\right\}. \label{eqn:rate1}
\end{align}
\end{theorem}
Note that if $S = \emptyset$, the above lower bound reduces to the secrecy capacity for the wiretap channel. Define
\begin{align*}
R_{\rm S-CSI-1} &= \max_{p(v|s)p(x|v,s)}\min\{I(V;Y|S) - I(V;Z|S) +H(S|Z), I(V;Y|S)\}, \\
R_{\rm S-CSI-2} &= \max_{p(v)p(x|v,s)}\min\{H(S|Z,V), I(V;Y|S)\}.
\end{align*}
Then, (\ref{eqn:rate1}) can be expressed as
\begin{align*}
C_{\rm S-CSI} \ge \max\{R_{\rm S-CSI-1}, R_{\rm S-CSI-2}\}.
\end{align*}
The proof of this theorem is detailed in Section~\ref{sect:4}. 

In~\cite{Liu--Chen2007}, the authors established the following lower bound for the noncausal case
\begin{align}
C_{\rm S-NCSI} &\ge \max_{p(u|s)p(x|u,s)} (I(U;Y,S)-\max\{I(U;Z), I(U;S)\}) \nonumber \\
& = \max_{p(u|s)p(x|u,s)}\min \left \{I(U;Y|S)-I(U;Z|S) + I(S;U|Z), I(U;Y|S)\right\}.\label{eqn:rate2}
\end{align} 
Clearly, $R_{\rm S-CSI-1}$ is at least as large as this lower bound. Hence, our lower bound (\ref{eqn:rate1}) is at least as large as this lower bound (\ref{eqn:rate2}). We now show that the lower bound (\ref{eqn:rate2}) is as large as $R_{\rm S-CSI-1}$. 

Fix $V\in [0:|\Vc|-1],$ $p(v|s),$ and $p(x|v,s)$ in $R_{\rm S-CSI-1}$. Let $U \in [0:|\Vc||\Sc|-1]$ in bound~(\ref{eqn:rate2}). Define the conditional probability mass functions: For $u = v + s|\Vc|$, let $p(u|s) = p(v|s), \, p(x|u,s) = p(x|v,s),$ and let $p(u|s) = p(x|u,s) =0$ otherwise. Under this mapping, it is easy to see that $H(S|Z,U) = 0$ and the other terms in (\ref{eqn:rate2}) reduce to those in $R_{\rm S-CSI-1}$.

We now show that our lower bound (\ref{eqn:rate1}) can be strictly larger than that for the noncausal case (\ref{eqn:rate2})). This is done via an example for which $R_{\rm S-CSI-2}> R_{\rm S-CSI-1}$. 

\begin{figure}[!ht]
\begin{center}
\psfrag{p}[c]{$0.1$}
\psfrag{0}[c]{$0$}
\psfrag{1}[c]{$1$}
\psfrag{x}[c]{$X$}
\psfrag{z}[c]{$Z$}
\psfrag{y}[c]{$Y$}
\psfrag{s}[c]{$S$}
\scalebox{1}{\includegraphics{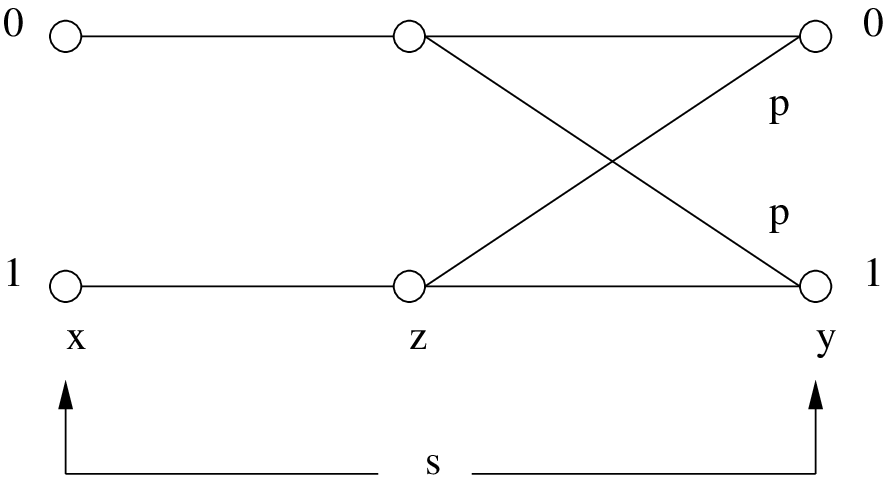}}
\caption{Example}\label{fig:0}
\end{center}
\end{figure}

Consider the channel in Figure~\ref{fig:0}, where $\Xc, \Yc, \Zc, \Sc \in \{0,1\}$ and $p(y,z|x,s) = p(y,z|x)$ with channel transition probabilities as defined in the Figure. The state $S$ is an i.i.d process that is observed by $X$ and $Y$ with $H(S) = 1 - H(0.1)$.

By setting $V = X$ independent of $S$ and $\P\{X = 1\} = \P\{X = 0\} = 0.5$ in $R_{\rm S-CSI-2}$, we obtain $R_{\rm S-CSI-2} = 1- H(0.1)$.

We now show that $R_{\rm S-CSI-1}$ is strictly smaller than $1 - H(0.1)$. First, note that
\begin{align*}
I(V;Y|S) &= H(Y|S) - H(Y|V,S) \\
& \le H(Y) - H(Y|X) \\
& = I(X;Y) \le 1-H(0.1). 
\end{align*}
However, for $R_{\rm S-CSI-1} \ge 1-H(0.1)$, we must have $I(V;Y|S) \ge 1- H(0.1)$. Hence, we must have $I(V;Y|S) = 1- H(0.1)$. Next, consider
\begin{align*}
I(V;Y|S) &= H(Y|S) - H(Y|V,S)\\
& \stackrel{(a)}{\le} 1 - H(Y|V,S) \\
& \stackrel{(b)}{\le} 1 - H(Y|V,S,X) \\
& = 1-H(0.1)
\end{align*}
Step $(a)$ holds with equality iff $p(y|s) = 0.5$ for all $y,s \in \{0,1\}$. From the structure of the channel, this implies that $p(x|s) = 0.5$ for all $x,s \in \{0,1\}$. Step $(b)$ holds with equality iff $H(Y|X,V,S) = H(Y|V,S)$, or equivalently $I(X;Y|V,S) = 0$. This implies that given $V,S$, $X$ and $Y$ are independent, $p(x,y|v,s) = p(x|v,s)p(y|v,s)$.
But since $p(x,y|v,s) = p(x|v,s)p(y|x)$, either (i) $p(x|v,s) = 0$ or (ii) $p(y|v,s) = p(y|x)$ must hold. Now, consider the pair $x=1, y=1$. Then, we must have either (i) $p(x=1|v,s) = 0$ or (ii) $p(y=1|v,s) = p(y=1|x=1) = 0.9$. In (i), $X$ is a function of $V$ and $S$. In (ii), we have 
\begin{align*}
p(y=1|v,s) &= p(x=1|v,s)p(y=1|x=1) + (1-p(x=1|v,s))p(y=1|x=0) \\
& = 0.9p(x=1|v,s) +0.1 -0.1p(x=1|v,s) \\
& = 0.8p(x=1|v,s) +0.1. 
\end{align*}
Using the fact that $p(y=1|v,s) = 0.9$, we have $0.8p(x=1|v,s) +0.1 = 0.9 \Rightarrow p(x=1|v,s) = 1$. This implies again that $X$ is a function of $V,S$. In both cases (i) and (ii), we see that $X$ is necessarily a function of $V$ and $S$, which implies that $Z =X$ is also a function of $V$ and $S$. Using the fact that $p(x|s) = p(z|s) =0.5$ for all $x,s$, we have
\begin{align*}
I(V;Z|S) &= H(Z|S) - H(Z|V,S) = H(X|S) =1.
\end{align*}
The first expression in $R_{\rm S-CSI-1}$ is then upper bounded by
\begin{align*}
I(V;Y|S) - I(V;Z|S) + H(S|Z)  & \le I(V;Y|S) - I(V;Z|S) + H(S) \\
& = 1-H(0.1) -1 + 1-H(0.1) \\
& = 1-2H(0.1) <1-H(0.1).
\end{align*}
This shows that $R_{\rm S-CSI-1} < R_{\rm S-CSI-2}$, which completes the example.

To illustrate the main ideas of the achievability proof of Theorem 1, we provide an outline for part of the proof of the rate expression $R_{\rm S-CSI-1}$. Using the functional representation lemma~\cite{Willems--Meulen1985}, we can show that it suffices to perform the maximization in $R_{\rm S-CSI-1}$ over $p(u), p(x|v,s)$, and functions $v(u,s)$. Thus, we prove achievability for the equivalent characterization of $R_{\rm S-CSI-1}$
\begin{align}
R_{\rm S-CSI-1} \ge \max_{p(u), v(u,s),p(x|v,s)}\min\{I(U;Y,S) - I(U;Z,S) + H(S|Z), I(U;Y,S)\}. \label{eqn:1}
\end{align}
We will outline the proof for the case where $I(U;Y,S) - I(U;Z,S)> 0$. Our coding scheme involves the transmission of $b-1$ independent messages over $b$ $n$-transmission blocks. We split the message $M_j$, $j \in [2:b]$, into two independent messages $M_{j0} \in [1:2^{nR_0}]$ and $M_{j1} \in [1:2^{nR_1}]$, where $R_0 + R_1 = R$. The codebook generation consists of two steps. The first step is the generation of the {\em message codebook}. We randomly generate $2^{nI(U;Y,S)}$ $u^n(l)$ sequences and partition  them into $2^{nR_0}$ equal size bins. The codewords in each bin are further partitioned into $2^{nR_1}$ equal size sub-bins $\Cc(m_0,m_1)$. The second step is to generate the key codebook. We randomly bin the set of state sequences $s^n$ into $2^{nR_K}$ bins $\Bc(k)$. The key $K_{j-1}$ used in block $j$ is the bin index of the state sequence $\Sv(j-1)$ in block $j-1$.

\begin{figure}[!ht]
\begin{center}
\psfrag{A}[r]{$M_j$}
\psfrag{m1}[c]{$M_{j1}$}
\psfrag{m0}[c]{$M_{j0}$}
\psfrag{m'}[c]{$M'_{j1}$}
\psfrag{k}[b]{$K_{j-1}$}
\psfrag{C}[c]{$U_i$}
\psfrag{E}[c]{$V_i$}
\psfrag{F}[l]{$X_i$}
\psfrag{P2}[c]{$p(s)$}
\psfrag{P3}[c]{$v(u,s)$}
\psfrag{P4}[c]{$p(x|v,s)$}
\psfrag{s1}[l]{$S_i$}
\psfrag{s2}[b]{$S_i$}
\includegraphics[width=0.85\linewidth]{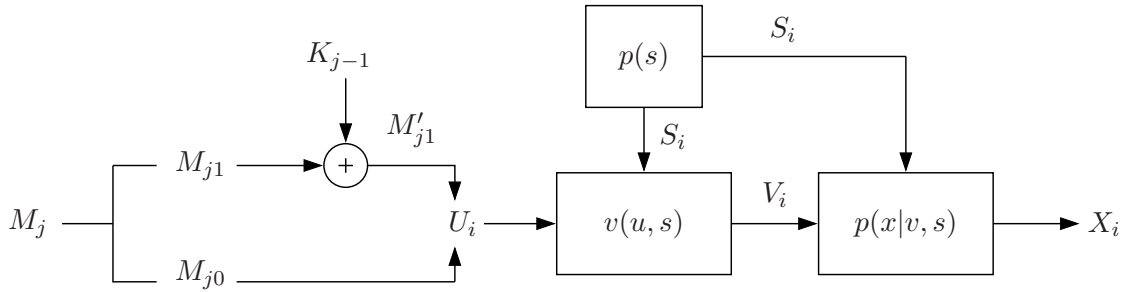}
\caption{Encoding in block $j$.}\label{fig:1}
\end{center}
\end{figure}

To send message $M_j$, $M_{j1}$ is encrypted with the key $K_{j-1}$ to obtain the index $M'_{j1}=M_{j1}\oplus K_{j-1}$. A codeword $u^n(L)$ is selected uniformly at random from sub-bin $\Cc(M_{j0}, M_{j1}\oplus K_{j-1})$ and transmitted using Shannon's strategy as depicted in Figure~\ref{fig:1}. The decoder uses joint typicality decoding together with its knowledge of the key to decode message $M_j$ at the end of block $j$. Finally, at the end of block $j$, the encoder and decoder declare the bin index $K_j$ of the state sequence $\sv(j)$ as the key to be used in block $j+1$. 
To show that the messages can be kept asymptotically secret from the eavesdropper, note that $M_{j0}$ is transmitted using Wyner wiretap coding. Hence, it can be kept secret from eavesdropper provided $I(U;Y,S) - I(U;Z,S) >0$. The key part of the proof is to show that the second part of the message $M_{j1}$, which is encrypted with the key $K_{j-1}$, can be kept secret from the eavesdropper. This involves showing that the eavesdropper has negligible information about $K_{j-1}$. However, the fact that $K_{j-1}$ is generated from the state sequence in block $j-1$ and used in block $j$ results in correlation between it and all received sequences at the eavesdropper from subsequent blocks. We show that the eavesdropper has negligible information about $K_{j-1}$ given all its received sequences provided $R_K < H(S|Z)$.

\subsection*{Upper Bound}
We establish the following upper bound on the secrecy capacity of the wiretap channel with noncausal state information available at both the encoder and decoder (which holds also for the causal case). 

\begin{theorem} \label{thm:2}
The following is an upper bound to the secrecy capacity of the DM-WTC with state noncausally available at the encoder and decoder 
\begin{align*}
C_{\rm S-NCSI} \le \min\left \{I(V_1;Y|U,S) - I(V_1;Z|U,S)+ H(S|Z,U), I(V_2;Y|S)\right\}.
\end{align*}
for some $U,$ $V_1$ and $V_2$ such that $p(u,v_1,v_2,x|s) = p(u|s)p(v_1|u,s)p(v_2|v_1,s)p(x|v_2,s).$
\end{theorem}
\medskip

The proof of this theorem is given in Section~\ref{sect:5}.
\subsection*{Secrecy Capacity Results}
\noindent 1. Following the lines of~\cite{Chen--Vinck2006}, we can show that Theorems~\ref{thm:1} and~\ref{thm:2} are tight for the following two special cases.
\begin{itemize}
\item[(i)] If there exists a $V^*$ such that $\max_{p(v|s)p(x|v,s)}(I(V;Y|S) - I(V;Z|S) + H(S|Z)) = I(V^*;Y|S) - I(V^*;Z|S) + H(S|Z)$ and $I(V^*;Y|S) - I(V^*;Z|S) + H(S|Z) \le I(V^*;Y|S)$, then the secrecy capacity is $C_{\rm S-CSI}=C_{\rm S-NCSI}  = I(V^*;Y|S) - I(V^*;Z|S) + H(S|Z)$.
\item[(ii)] If there exists a $V'$ such that $\max_{p(v|s)p(x|v,s)}I(V;Y|S)= I(V';Y|S)$ and $I(V';Y|S) \le I(V';Y|S) - I(V';Z|S) + H(S|Z)$, then the secrecy capacity is $C_{\rm S-CSI}=C_{\rm S-NCSI} = I(V';Y|S)$.
\end{itemize}

\noindent 2. We show that Theorems~\ref{thm:1} and~\ref{thm:2} are also tight when $I(U;Y|S)\ge I(U;Z|S)$ for $U$ such that $(U,S) \to (X,S) \to (Y,Z)$ form a Markov chain, i.e., when $Y$ is {\em less noisy} than $Z$ for every state $s\in\Sc$~\cite{Korner--Marton}.  

\begin{theorem} \label{thm:3}
The secrecy capacity for the DM-WTC with the state information available causally or noncausally at the encoder and decoder when $Y$ is {less noisy} than $Z$ is
\begin{align*}
C_{\rm S-CSI}& =C_{\rm S-NCSI} = \max_{p(x|s)}\min\{I(X;Y|S)- I(X;Z|S) + H(S|Z), I(X;Y|S)\}.
\end{align*} 
\end{theorem}
Consider the special case when $p(y,z|x,s) = p(y,z|x)$ and $Z$ is a degraded version of $Y$, then Theorem~\ref{thm:3} specializes to the secrecy capacity for the wiretap channel with a key~\cite{Yamamoto1997}
\begin{align*}
& C_{\rm S-CSI}=C_{\rm S-NCSI} \\ 
& = \max_{p(x)} \min\{I(X;Y) - I(X;Z) + H(S), I(X;Y)\}.
\end{align*}

Achievability for Theorem~\ref{thm:3} follows directly from Theorem~\ref{thm:1} by setting $V= X$ and observing that the expression reduces to $R_{\rm S-CSI-1}$ since $Y$ is less noisy than $Z$. To establish the converse, we use the less noisy assumption to strengthen the first inequality in Theorem~\ref{thm:2} as follows
\begin{align*}
I(V_1;Y|U,S) - I(V_1;Z|U,S) + H(S|Z,U) &\le I(V_1;Y|U,S) - I(V_1;Z|U,S) + H(S|Z) \\
&\stackrel{(a)}{\le} I(V_1;Y|S) - I(V_1;Z|S) + H(S|Z) \\
& \stackrel{(b)}{\le} I(X;Y|S) - I(X;Z|S) + H(S|Z),
\end{align*}
where $(a), (b)$ follow from the less noisy assumption. The proof of the second inequality follows by the data processing inequality: $I(V_2;Y|S) \le I(X;Y|S)$.

\noindent 3. Next, consider the case where $p(y,z|x,s) = p(y,z|x)$ and the eavesdropper $Z$ is {\em less noisy}~\cite{Korner--Marton} than $Y.$ That is, $I(U;Z) \ge I(U;Y)$ for every $U$ such that $U \to X \to (Y,Z).$ Then, the capacity of this special class of channels is 
\begin{align*}
C_{\rm S-CSI} = C_{\rm S-NCSI}= \max_{p(x)}\min\{H(S), I(X;Y)\}.
\end{align*}
Achievability follows by setting $V=X$ independent of $S$. The converse follows from Theorem~\ref{thm:2} and the observation that since $Z$ is less noisy than $Y$ and $p(y,z|x,s) = p(y,z|x),$
\begin{align*}
I(V_1;Y|U,S) - I(V_1;Z|U,S) + H(S|Z, U) &\le H(S|Z,U) \\
& \le H(S),
\end{align*}
and $I(V_2;Y|S) \le I(X;Y)$.


\section{Proof of Theorem~\ref{thm:1}} \label{sect:4}
We will prove the achievability of $R_{\rm S-CSI-1}$ and $R_{\rm S-CSI-2}$ separately. For $R_{\rm S-CSI-1}$, we will prove the equivalent expression stated in equation~\ref{eqn:1}.

The proof of achievability for $R_{\rm S-CSI-1}$ is split into two cases (Cases 1 and 2) while $R_{\rm S-CSI-2}$ is proved in Case 3.

\subsection*{Case 1: $R_{\rm S-CSI-1}$ with $I(U;Y,S) > I(U;Z,S)$}

\subsubsection*{Codebook generation}
Split message $M_j$ into two independent messages $M_{j0}\in [1:2^{nR_0}]$ and $M_{j1} \in [1:2^{nR_1}]$, thus $R=R_0+R_1$. Let $\Rt \ge R$. The codebook generation consists of two steps. 

{\em Message codeword generation}: We randomly and independently generate $2^{n\Rt}$ sequences $u^n(l)$, $l \in [1:2^{n\Rt}]$, each according to $\prod_{i=1}^n p(u_i)$ and partition them into $2^{nR_0}$ equal-size bins $\Cc(m_0)$, $m_0 \in [1:2^{nR_0}]$. We further partition the sequences within each bin $\Cc(m_0)$ into $2^{nR_K}$ equal size sub-bins, $\Cc(m_0, m_1)$, $m_1  \in [1:2^{nR_1}]$.

{\em Key codebook generation}: We randomly and uniformly partition the set of $s^n$ sequences into $2^{nR_K}$ bins $\Bc(k)$, $k\in [1:2^{nR_K}]$. 

Both codebooks are revealed to all parties.
\medskip

\subsubsection*{Encoding} 
We send $b-1$ messages over $b$ $n$-transmission blocks. In the first block, we randomly select a sequence $u^n(L) \in \Cc(m_{10},m'_{11})$. The encoder then computes $v_i= v(u_i(L), s_i)$ and transmits a randomly generated symbol $X_i \sim p(x_i|s_i,v_i)$ for $i \in [1:n]$.
At the end of the first block, the encoder and decoder declare $k_1 \in [1:2^{nR_K}]$ such that $\sv(1) \in \Bc(k_1)$ as the key to be used in block 2. 

Encoding in block $j \in [2:b]$ proceeds as follows. To send message $m_j=(m_{j0},m_{j1})$ and given key $k_{j-1}$, the encoder computes $m'_{j1} = m_{j1}\oplus k_{j-1}$. To ensure secrecy, we must have $R_1 \le R_K$~\cite{Shannon1949}. The encoder then randomly selects a sequence $u^n(L) \in \Cc(m_{j0}, m'_{j1})$. It then computes $v_i= v(u_i(L), s_i)$ and transmits a randomly generated symbol $X_i \sim p(x_i|s_i,v_i)$ for $i \in [(j-1)n+1:jn]$.
 
\subsubsection*{Decoding and analysis of the probability of error} At the end of block $j$, the decoder declares that $\lh$ is sent if it is the unique index such that $(u^n(\lh), \Yv(j), \Sv(j))\in \aep$, otherwise it declares an error. It then finds the indices $(\mh_{j0},\mh'_{j1})$ such that $u^n(l) \in \Cc(\mh_{j0},\mh'_{j1})$. Finally, it recovers $\mh_{j1}$ by computing $\mh_{j1} = \mh_{j1}'\oplus k_{j-1}$.

To analyze the probability of error, let $\e'' > \e' > \e>0$ and define the following events for every $j\in [2:b]$:
\begin{align*}
\Ec(j) &= \{\Mh_j \neq M_j\}, \\
\Ec_1(j) & = \{(U^n(L), \Sv(j)) \notin \mathcal{T}^n_{\e'}\}, \\
\Ec_2(j) & = \{(U^n(L),\Sv(j), \Yv(j)) \notin \mathcal{T}^n_{\e''}\}, \\
\Ec_3(j) & = \{(U^n(\lh),\Sv(j), \Yv(j)) \in \mathcal{T}^n_{\e''} \text{ for some } \lh \neq L\}.
\end{align*}
The probability of error is upper bounded as 
\begin{align*}
\P(\Ec) & = \P\{\cup_{j=2}^b \Ec(j)\}  \le \sum_{j=2}^b \P(\Ec(j)).
\end{align*}
Each probability of error term can be upper bounded as 
\begin{align*}
\P(\Ec(j)) &\le \P(\Ec_1(j)) + \P(\Ec_2(j)\cap\Ec^c_1(j)) + \P(\Ec_3(j)\cap\Ec^c_2(j)). 
\end{align*}
Now, $\P(\Ec_1(j))\to 0$ as $n\to \infty$ by Law of Large Numbers (LLN) since $\P\{(U^n(L) \in \aep)\} \to 1$ as $n \to \infty$ and $\Sv(j)\sim\prod_{i=1}^n p(s_i) = \prod_{i=1}^n p(s_i|u_i)$ by independence. The term $\P(\Ec_2(j)\cap\Ec^c_1(j))\to 0$ as $n \to \infty$ by LLN since $(U^n(L), \Sv(j) \in \mathcal{T}_{e'}^n$ and $Y^n\sim\prod_{i=1}^n p(y_i|u_i, s_i)$. For the last term, consider 
{\allowdisplaybreaks
\begin{align*}
\P(\Ec_3\cap\Ec^c_2(j)) &\le \sum_{l}p(l) \sum_{\lh\neq l} \P\{(U^n(\lh),\Sv(j), \Yv(j)) \in \mathcal{T}^n_{\e''}|\Ec_2^c(j), L=l\} \\
&\stackrel{(a)}{\le} \sum_{\lh \neq l} 2^{-n(I(U;Y,S) - \d(\e''))} \le 2^{n(\Rt - I(U;Y,S) + \d(\e''))},
\end{align*}
where $(a)$ follows from: (i) $L$ is independent of the transmission codebook sequences $U^n$ and the current state sequence $\Sv(j)$; and (ii) the conditional joint typicality lemma~\cite[Lecture 2]{El-Gamal--Kim2010}. Hence, $\P(\Ec_3\cap\Ec^c_2(j))\to 0$ as $n\to \infty$ if $\Rt < I(U;Y,S) - \d(\e'')$.}
\medskip

\subsubsection*{Analysis of the information leakage rate} We use $\Zv^j$ to denote the eavesdropper's received sequence from blocks 1 to $j$ and $\Zv(j)$ to denote the received sequence in block $j$. We will need the following two results. 

\begin{proposition} \label{prop1}
If $R_{K} < H(S|Z) - 4\d(\e)$ and $\Rt \ge I(U;Z,S)$, then the following holds for every $j\in [1:b]$.
\begin{enumerate}
\item $H(K_{j}|\Cc) \ge n(R_K - \d(\e))$.
\item $I(K_j; \Zv(j)|\Cc) \le 2n\d(\e)$.  
\item $I(K_j;\Zv^{j}|\Cc) \le n\d'(\e)$, where $\d(\e)\to 0$ and $\d'(\e)\to 0$ as $\e \to 0$. 
\end{enumerate}
\end{proposition}
The proof of this proposition is given in Appendix~\ref{appen1}. 
\medskip

\begin{lemma}{~\cite{Chia--El-Gamal_a}} \label{lem1}
Let $(U,V,Z) \sim p(u,v,z)$, $\bar{R} \ge 0$ and $\e >0$. Let $U^n$ be a random sequence distributed according to $\prod_{i=1}^np(u_i)$. Let $V^n(l)$, $l \in [1:2^{n\bar{R}}]$, be a set of random sequences that are conditionally independent given $U^n$ and each distributed according to $\prod_{i=1}^n p(v_i|u_i)$. Let $L$ be a random index with an arbitrary distribution over $[1:2^{n\bar{R}}]$ independent of $(U^n, V^n(l)), l \in [1:2^{n\bar{R}}]$. Then, if $\P\{(U^n,V^n(L), Z^n)\in \aep\}\to 1$ as $n\to \infty$ and $\bar{R} \geq I(V;Z|U)$, there exists a $\d(\e)>0$, where $\d(\e) \to 0$ as $\e \to 0$, such that
$H(L|Z^n,U^n) \leq n(\bar{R} - I(V;Z|U)) +n\d(\e)$.
\end{lemma} 
\medskip

We are now ready to upper bound the leakage rate averaged over codes. Consider 
\begin{align*}
I(M_2,M_3, \ldots, M_b;\Zv^b|\Cc)& = \sum_{j=2}^b I(M_{j};\Zv^b|\Cc, M_{j+1}^b) \\
& \stackrel{(a)}{\le} \sum_{j=2}^b I(M_{j};\Zv^b|\Cc, \Sv(j), M_{j+1}^b) \\
& \stackrel{(b)}{=} \sum_{j=2}^b I(M_{j};\Zv^j|\Cc, \Sv(j)),
\end{align*}
where $(a)$ follows by the independence of $M_{j}$ and $(\Sv(j), M_{j+1}^b)$, and $(b)$ follows by the Markov Chain relation $(\Zv_{j+1}^b, M_{j+1}^b,\Cc) \to (\Zv^j,\Sv(j),\Cc) \to (M_j, \Cc)$. Hence, it suffices to upper bound each individual term $I(M_{j};\Zv^j|\Cc, \Sv(j))$. Consider
\begin{align*}
I(M_{j};\Zv^j|\Cc, \Sv(j)) & = I(M_{j0}, M_{j1};\Zv^j|\Cc, \Sv(j)) \\
& = I(M_{j0}, M_{j1};\Zv^{j-1}|\Cc, \Sv(j)) + I(M_{j0}, M_{j1};\Zv(j)|\Cc, \Sv(j), \Zv^{j-1}).
\end{align*}
Note that the first term is equal to zero by the independence of $M_j$ and past transmissions, the codebook, and state sequence. For the second term, we have
\begin{align*}
I(M_{j0}, M_{j1};\Zv(j)|\Cc, \Sv(j), \Zv^{j-1}) &= I(M_{j0};\Zv(j)|\Cc, \Sv(j), \Zv^{j-1}) + I(M_{j1};\Zv(j)|\Cc, M_{j0}, \Sv(j), \Zv^{j-1}).
\end{align*}
We now bound the each term separately. Consider the first term {\allowdisplaybreaks
\begin{align*}
I(M_{j0};\Zv(j)|\Cc, \Sv(j), \Zv^{j-1}) & = I(M_{j0}, L; \Zv(j)|\Cc, \Sv(j), \Zv^{j-1}) -  I(L; \Zv(j)|\Cc, M_{j0}, \Sv(j), \Zv^{j-1}) \\
& \le I(U^n;\Zv(j)|\Cc, \Sv(j), \Zv^{j-1}) -H(L|\Cc, M_{j0}, \Sv(j), \Zv^{j-1}) \\
& \qquad  + H(L|\Zv(j), M_{j0}, \Sv(j)) \\
& \le \sum_{i=1}^n(H(\Zv_i(j)|\Cc, \Sv_i(j)) - H(\Zv_i(j)|\Cc, U_i, \Sv_i(j))) \\ 
& \qquad -H(L|\Cc, M_{j0}, \Sv(j), \Zv^{j-1})+ H(L|\Zv(j), M_{j0}, \Sv(j)) \\
& \stackrel{(a)}{\le} nI(U;Z|S)-H(L|\Cc, M_{j0}, \Sv(j), \Zv^{j-1}) + H(L|\Zv(j), M_{j0}, \Sv(j)) \\
& \stackrel{(b)}{\le} nI(U;Z|S) -H(L|\Cc, M_{j0}, \Sv(j), \Zv^{j-1}) \\ & \quad + n(\Rt - R_0 - I(U;Z,S) + \d(\e)) \\
& \stackrel{(c)}{=} n(\Rt-R_0) - H(L|\Cc, M_{j0}, \Sv(j), \Zv^{j-1}) +n\d(\e) \\
& = n(\Rt-R_0) - H(M_{j1}\oplus K_{j-1}|\Cc, M_{j0}, \Sv(j), \Zv^{j-1}) \\
& \qquad - H(L|\Cc, M_{j0}, \Sv(j), \Zv^{j-1}, M_{j1}\oplus K_{j-1}) +n\d(\e) \\
& \le n(\Rt-R_0) - H(M_{j1}\oplus K_{j-1}|\Cc, M_{j0}, \Sv(j), K_{j-1}, \Zv^{j-1}) \\
& \qquad - n(\Rt-R_0 - R_K) +n\d(\e) \\
& \stackrel{(d)}{=} nR_K - H(M_{j1}\oplus K_{j-1}|\Cc, M_{j0}, \Sv(j), K_{j-1}) + n\d(\e)\\
& = nR_K - H(M_{j1}|\Cc, M_{j0}, \Sv(j), K_{j-1}) + n\d(\e) = n\d(\e),
\end{align*}
where $(a)$ follows from the fact that $H(\Zv_i(j)|\Cc, \Sv_i(j)) \le H(\Zv_i(j)|\Sv_i(j))= H(Z|S)$ and $H(\Zv_i(j)|\Cc, U_i, \Sv_i(j)) = H(Z|U,S)$. Step $(b)$ follows by Lemma 1 which requires that (i) $\P\{(U^n(L),\Sv(j), \Zv(j)) \in \aep\}\to 1$ as $n\to \infty$, and (ii) $\Rt - R_0 \ge I(U;Z,S)$; where (i) can be shown using the same steps as in the analysis of probability of error. Step $(c)$ follows by the independence of $U$ and $S$. Step $(d)$ follows from the Markov Chain relation $(\Zv^{j-1}, M_{j0}, \Sv(j)) \to (K_{j-1}, M_{j0}, \Sv(j)) \to (M_{j1}\oplus K_{j-1}, M_{j0}, \Sv(j))$. The last step follows by the fact that $M_{j1}$ is independent of $(\Cc, M_{j0}, \Sv(j), K_{j-1})$ and uniformly distributed over $[1:2^{nR_K}]$.

Next, consider the second term 
\begin{align*}
I(M_{j1};\Zv(j)|\Cc, M_{j0}, \Sv(j), \Zv^{j-1}) &\le I(M_{j1},L;\Zv(j)|\Cc, M_{j0}, \Sv(j), \Zv^{j-1}) \\
& \qquad - I(L;\Zv(j)|\Cc, M_{j0}, M_{j1}, \Sv(j), \Zv^{j-1}) \\
& \le I(U^n;\Zv(j)|\Cc, M_{j0}, \Sv(j), \Zv^{j-1}) - H(L|\Cc, M_{j0}, M_{j1}, \Sv(j), \Zv^{j-1}) \\
& \qquad + H(L|\Cc, M_{j0}, M_{j1}, \Sv(j), \Zv^{j}) \\
& \stackrel{(a)}{\le} nI(U;Z|S) - H(L|\Cc, M_{j0}, M_{j1}, \Sv(j), \Zv^{j-1}) \\
& \qquad + H(L|\Cc, M_{j0}, M_{j1}, \Sv(j), \Zv^{j}) \\
& \le nI(U;Z|S) - H(L|\Cc, M_{j0}, M_{j1}, \Sv(j), \Zv^{j-1}) + H(L|M_{j0}, \Sv(j), \Zv(j)) \\
& \stackrel{(b)}{\le} nI(U;Z|S) - H(L|\Cc, M_{j0}, M_{j1}, \Sv(j), \Zv^{j-1}) \\
& \qquad + n(\Rt-R_0) - nI(U;Z,S)+n\d(\e) \\
& = n(\Rt-R_0) - H(L|\Cc, M_{j0}, M_{j1}, \Sv(j), \Zv^{j-1}) + n\d(\e),
\end{align*}
where $(a)$ follows from the same steps used in bounding $I(M_{j0};\Zv(j)|\Cc, \Sv(j), \Zv^{j-1})$; $(b)$ follows from Lemma 1. Next consider
\begin{align*}
H(L|\Cc, M_{j0}, M_{j1}, \Sv(j), \Zv^{j-1}) &= H(M_{j1}\oplus K_{j-1}|\Cc, M_{j0}, M_{j1}, \Sv(j), \Zv^{j-1}) \\
& \quad + H(L|\Cc, M_{j0}, M_{j1}, M_{j1}\oplus K_{j-1}, \Sv(j), \Zv^{j-1}) \\
& = H(K_{j-1}|\Cc, M_{j0}, M_{j1}, \Sv(j), \Zv^{j-1}) + n(\Rt - R_0 - R_K) \\
& = H(K_{j-1}|\Cc, \Zv^{j-1}) + n(\Rt - R_0 - R_K).
\end{align*}}
From Proposition~\ref{prop1}, $H(K_{j-1}|\Cc, \Zv^{j-1}) \ge n(R_K  -\d(\e)-\d'(\e))$, which implies that
\begin{align*}
I(M_{j1};\Zv(j)|\Cc, M_{j0}, \Sv(j), \Zv^{j-1}) &\le n(\d'(\e) + 2\d(\e)).
\end{align*}
This completes the analysis of information leakage rate.

\subsubsection*{Rate analysis} 
From the analysis of probability of error and information leakage rate, we see that the rate constraints are 
\begin{align*}
\Rt &< I(U;Y,S) -\d(\e), \\
\Rt -R_0 &\ge I(U;Z,S), \\
R_K &< H(S|Z) -4\d(\e), \\
R_0+R_1 &\le \Rt, \\
R_1 &\le R_K, \\
R &= R_0 + R_1.
\end{align*}
Using Fourier-Motzkin elimination (see for e.g. Lecture 6 of~\cite{El-Gamal--Kim2010}), we obtain
\begin{align*}
R &< \max_{p(u), v(u,s), x(u,s)}\min \{I(U;Y,S) - I(U;Z,S) + H(S|Z), I(U;Y,S)\} \\
& \stackrel{(a)}{=} \max_{p(u), v(u,s), p(x|s,v)}\min \{I(V;Y|S) - I(V;Z|S) + H(S|Z), I(V;Y|S)\},
\end{align*}
where $(a)$ follows by the independence of $U$ and $S$ and the fact that $V$ is a function of $U$ and $S$. 

\subsection*{Case 2: $R_{\rm S-CSI-1}$ with $I(U;Y,S) \le I(U;Z,S)$}

Under this condition, the decoder cannot rely on the wiretap channel to send a confidential message. Therefore, only the key is used to encrypt the message and transmit it securely. Note that we only need to consider the case where $H(S|Z) - (I(U;Z,S)- I(U;Y,S)) >0$. 

\subsubsection*{Codebook generation} Codebook generation again consists of two steps.

{\em Message codebook generation}: Let $\Rt \ge R_d$ and $R \le \Rt-R_d$. Randomly and independently  generate $2^{n\Rt}$ sequences $u^n(l)$, $l \in [1:2^{n\Rt}]$, each according to $\prod_{i=1}^n p(u_i)$ and partition them into $2^{nR_d}$ equal-size bins $\Cc(m_d)$, $m_d \in [1:2^{nR_d}]$. We further partition the set of sequences in each  bin $\Cc(m_d)$ into sub-bins, $\Cc(m_d, m)$, $m \in [1:2^{nR}]$.

{\em Key codebook generation}: We randomly bin the set of $s^n \in \Sc^n$ sequences into $2^{nR_K}$ bins $\Bc(k)$, $k\in [1:2^{nR_K}]$. 

\subsubsection*{Encoding} We send $b-1$ messages over $b$ $n$-transmission blocks. In the first block, we randomly select a $u^n(L)$ sequence. The encoder then computes $v_i= v(u_i(L), s_i)$, $i \in [1:n]$, and transmits a randomly generates sequence $X^n$ according to $\prod_{i=1}^n p(x_i|s_i,v_i)$. At the end of the first block, the encoder and decoder declare $k_1 \in [1:2^{nR_K}]$ such that $\sv(1) \in \Bc(k_1)$ as the key to be used in block 2. 

Encoding in block $j \in [2:b]$ is as follows. We split the key $k_{j-1}$ into two independent parts, $K_{j-1,d}$ and $K_{j-1, m}$ at rates $R_d$ and $R$, respectively. To send message $m_j$, the encoder computes $m' = m_j\oplus k_{(j-1)m}$. This requires that $R_K \ge R + R_d$. The encoder then randomly selects a sequence $u^n(L) \in \Cc(k_{(j-1)d}, m')$. At time  $i \in [(j-1)n+1:jn]$, it computes $v_i= v(u_i(L), s_i)$, and transmits a randomly generated symbol $X_i$ according to $p(x_i|s_i,v_i)$.

\subsubsection*{Decoding and analysis of the probability of error} At the end of block $j$, the decoder declares that $\lh$ is sent if it is the unique index such that $(u^n(\lh), \Yv(j), \Sv(j))\in \aep$ and $u^n(\lh) \in \Cc(k_{(j-1)d})$. Otherwise, it declares an error. It then finds the index $\mh'$ such that $u^n(\lh) \in \Cc(k_{(j-1)d}, \mh')$. Finally, it recovers $\mh_j$ by computing $\mh_j = \mh' \oplus k_{(j-1)m}$. Following similar steps to the analysis for Case 1, it can be shown that $\P_e \to 0$ as $n \to \infty$ if $\Rt - R_d < I(U;Y,S) - \d(\e)$.

\subsubsection*{Analysis of the information leakage rate} Following the same steps as for Case 1, we can show that it suffices to upper bound the terms $I(M_j;\Zv(j)|\Cc, \Sv(j), \Zv^{j-1})$ for $j \in [2:b]$. Consider
\begin{align*}
I(M_j;\Zv(j)|\Cc, \Sv(j), \Zv^{j-1}) &= H(M_j) - H(M_j|\Cc, \Sv(j), \Zv^{j}) \\
&\le H(M_j) - H(M_j|\Cc, \Sv(j), K_{(j-1)d}, M_{j}\oplus K_{(j-1)m},\Zv^{j}) \\
&= H(M_j) - H(M_j|\Cc, K_{(j-1)d}, M_{j}\oplus K_{(j-1)m},\Zv^{j-1}) \\
&= H(M_j) - H(M_j|\Cc, \Zv^{j-1}, K_{(j-1)d})- H(M_{j}\oplus K_{(j-1)m}|\Cc, \Zv^{j-1},K_{(j-1)d},M_j)\\
&\qquad +H(M_{j}\oplus K_{(j-1)m}|\Cc, \Zv^{j-1}, K_{(j-1)d}) \\
& = nR -H(M_j) +H(M_{j}\oplus K_{(j-1)m}|\Cc, \Zv^{j-1}, K_{(j-1)d}) \\
& \quad - H(M_{j}\oplus K_{(j-1)m}|\Cc, \Zv^{j-1},K_{(j-1)d},M_j)\\
& \le nR - H(K_{(j-1)m}|\Cc,\Zv^{j-1}, K_{(j-1)d}).
\end{align*}
Thus, showing that
\begin{align}
I(K_{(j-1)m};\Zv^{j-1}|\Cc, K_{(j-1)d}) &\le n\d'(\e), \label{case2_1}\\
H(K_{(j-1)m}|\Cc, K_{(j-1)d}) &\ge n(R_K-R_d - \d(\e)) \label{case2_2}
\end{align}
implies 
\begin{align*}
I(M_j;\Zv(j)|\Cc, \Sv(j), \Zv^{j-1}) \le nR - n(R_K - R_d) + n(\d'(\e)+\d(\e)).
\end{align*}
Hence, the rate of information leakage approaches zero as $n\to \infty$ if $R \le R_K - R_d$. To prove~(\ref{case2_1}) and~(\ref{case2_2}), we need the following Proposition.

\begin{proposition} \label{prop2}
If $\Rt \ge I(U;Z,S)$ and $R_K < H(S|Z) -4\d(\e)$, then for all $j\in [1:b]$,
\begin{enumerate}
\item $H(K_{j}|\Cc) \ge n(R_K - \d(\e))$.
\item $I(K_j; \Zv(j)|\Cc) \le 3n\d(\e)$.
\item $I(K_{j};\Zv^{j}|\Cc) \le n\d'(\e)$, where $\d(\e)\to 0$ and $\d'(\e)\to 0$ as $\e \to 0$.
\end{enumerate}
\end{proposition}
The proof of this Proposition is given in Appendix~\ref{appen2}. 

Part 3 of Proposition~\ref{prop2} implies~(\ref{case2_1}), since 
\begin{align*}
I(K_{j-1};\Zv^{j-1}|\Cc) &= I(K_{(j-1)d}, K_{(j-1)m};\Zv^{j-1}|\Cc) \\
& = I(K_{(j-1)d};\Zv^{j-1}|\Cc) + I(K_{(j-1)m};\Zv^{j-1}|\Cc, K_{(j-1)d}).
\end{align*}

Part 1 of Proposition 2 implies~(\ref{case2_2}), since $H(K_{(j-1)}|\Cc) = H(K_{(j-1)m}, K_{(j-1)d}|\Cc)\ge n(R_K - \d(\e))$, which implies that $H(K_{(j-1)m}|\Cc, K_{(j-1)d}) \ge n(R_K-R_d - \d(\e))$.

\subsubsection*{Rate analysis} The following rate constraints are necessary for Case 2.{\allowdisplaybreaks
\begin{align*}
\Rt &\ge I(U;Z,S), \\
\Rt - R_d & < I(U;Y,S) - \d(\e), \\
R &\le \Rt - R_d, \\
R_K &< H(S|Z) -4\d(\e), \\
R &\le R_K -R_d.
\end{align*}
}
Using Fourier Motzkin elimination, we obtain
\begin{align*}
R &< \max_{p(u), v(u,s)}\min \{I(U;Y,S) - I(U;Z,S) + H(S|Z), I(U;Y,S)\} \\
& = \max_{p(u), v(u,s), p(x|s,v)}\min \{I(V;Y|S) - I(V;Z|S) + H(S|Z), I(V;Y|S)\}.
\end{align*}

\subsection*{Case 3: $R_{\rm S-CSI-2}$}
For $R_{\rm S-CSI-2}$, the key generated in a block is used purely to encrypt the message in the following block. This implies that there is a possibility that the eavesdropper can decode the codeword transmitted in the current block, which reduces the key rate that can be generated at the current block. This is compensated for by the fact that the entire key is used for message transmission. The codebook generation, encoding and analysis of probability of error and equivocation are therefore similar to that in Case 2.

\subsubsection*{Codebook generation} Codebook generation again consists of two steps. 

{\em Message codebook generation}: Randomly and independently  generate $2^{nR}$ sequences $v^n(l)$, $l \in [1:2^{nR}]$, each according to $\prod_{i=1}^n p(v_i).$

{\em Key codebook generation}: Set $R_K = R.$ We randomly bin the set of $s^n \in \Sc^n$ sequences into $2^{nR_K}$ bins $\Bc(k)$, $k\in [1:2^{nR_K}]$. 
\medskip

\subsubsection*{Encoding} We send $b-1$ messages over $b$ $n$-transmission blocks. In the first block, we randomly select a $v^n(L)$ sequence. The encoder then transmits a randomly generated sequence $X^n$ according to $\prod_{i=1}^n p(x_i|s_i,v_i)$. At the end of the first block, the encoder and decoder declare $k_1 \in [1:2^{nR_K}]$ such that $\sv(1) \in \Bc(k_1)$ as the key to be used in block 2. 

Encoding in block $j \in [2:b]$ is as follows. To send message $m_j$, the encoder computes $m' = m_j\oplus k_{j-1}$. The encoder then selects the sequence $v^n(m')$. At time  $i \in [(j-1)n+1:jn]$, it transmits a randomly generated symbol $X_i$ according to $p(x_i|s_i,v_i)$.
\medskip

\subsubsection*{Decoding and analysis of the probability of error} At the end of block $j$, the decoder declares that $\mh'$ is sent if it is the unique index such that $(v^n(\mh'), \Yv(j), \Sv(j))\in \aep$. Otherwise, it declares an error. It then recovers $\mh_j$ by computing $\mh_j = \mh' \oplus k_{j-1}$. Following similar steps to the analysis for Case 1, it can be shown that $\P_e \to 0$ as $n \to \infty$ if $R < I(V;Y,S) - \d(\e)$.
\medskip

\subsubsection*{Analysis of the information leakage rate} Following the same steps as for Case 1, we can show that it suffices to upper bound the terms $I(M_j;\Zv(j)|\Cc, \Sv(j), \Zv^{j-1})$ for $j \in [2:b]$. Consider
\begin{align*}
I(M_j;\Zv(j)|\Cc, \Sv(j), \Zv^{j-1}) &= H(M_j) - H(M_j|\Cc, \Sv(j), \Zv^{j}) \\
&\le H(M_j) - H(M_j|\Cc, \Sv(j), M_{j}\oplus K_{j-1},\Zv^{j}) \\
& = H(M_j) - H(M_j|\Cc, M_{j}\oplus K_{j-1},\Zv^{j-1}) \\
& = H(M_j) - H(M_{j}\oplus K_{j-1}, M_{j}|\Cc, \Zv^{j-1}) + H(M_j\oplus K_{j-1}|\Cc, \Zv^{j-1}) \\
& \le nR - H(M_{j}|\Cc, \Zv^{j-1}) - H(M_j\oplus K_{j-1}|\Cc,\Zv^{j-1}, M_j) + nR \\
& = nR - H(K_{j-1}|\Cc,\Zv^{j-1}).
\end{align*}
Thus, showing that
\begin{align}
I(K_{j-1};\Zv^{j-1}|\Cc) &\le n\d'(\e), \label{case3_1}\\
H(K_{j-1}|\Cc) &\ge n(R_K - \d(\e)) \label{case3_2}
\end{align}
implies 
\begin{align*}
I(M_j;\Zv(j)|\Cc, \Sv(j), \Zv^{j-1}) \le n(\d'(\e)+\d(\e)).
\end{align*}
To prove~(\ref{case3_1}) and~(\ref{case3_2}), we will use the following Proposition

\begin{proposition} \label{prop3}
If $R_K < H(S|Z, V) -4\d(\e)$, then for all $j\in [1:b]$,
\begin{enumerate}
\item $H(K_{j}|\Cc) \ge n(R_K - \d(\e))$.
\item $I(K_j; \Zv(j)|\Cc) \le 3n\d(\e)$.
\item $I(K_{j};\Zv^{j}|\Cc) \le n\d'(\e)$, where $\d(\e)\to 0$ and $\d'(\e)\to 0$ as $\e \to 0$.
\end{enumerate}
\end{proposition}
The proof of this Proposition is given in Appendix~\ref{appen3}. It is clear that equations (\ref{case3_1}) and (\ref{case3_2}) are implied by Proposition~\ref{prop3}, which completes the analysis of information leakage rate.
\medskip

\subsubsection*{Rate analysis} The following rate constraints are necessary for Case 3.{\allowdisplaybreaks
\begin{align*}
R &= R_K, \\
R &< I(V;Y,S) - \d(\e), \\
R_K &< H(S|Z,V) -4\d(\e).
\end{align*}
}
These constraints imply the achievability of 
\begin{align*}
R &< \max_{p(v)p(x|s,v)}\min \{H(S|Z,V), I(V;Y|S)\}.
\end{align*}

\section{Proof of Theorem~\ref{thm:2}} \label{sect:5}
For any sequence of codes with probability of error and leakage rate that approach zero as $n \to \infty$, consider  {\allowdisplaybreaks
\begin{align*}
nR & = H(M) \stackrel{(a)}{\le} I(M;Y^n,S^n) + n\e_n \\
& \stackrel{(b)}{\le} I(M;Y^n,S^n) - I(M;Z^n) + 2n\e_n \\
& = \sum_{i=1}^n(I(M;Y_i,S_i|Y_{i+1}^n, S_{i+1}^n) - I(M;Z_i|Z^{i-1})) + 2n\e_n\\
& \stackrel{(c)}{=}\sum_{i=1}^n(I(M, Z^{i-1};Y_i,S_i|Y_{i+1}^n, S_{i+1}^n) - I(M, Y_{i+1}^n, S_{i+1}^n;Z_i|Z^{i-1})) + 2n\e_n\\
& \stackrel{(d)}{=}\sum_{i=1}^n(I(M;Y_i,S_i|Y_{i+1}^n, S_{i+1}^n, Z^{i-1}) - I(M;Z_i|Y_{i+1}^n, S_{i+1}^n, Z^{i-1})) + 2n\e_n\\
& \stackrel{(e)}{=}\sum_{i=1}^n(I(V_{1i};Y_i,S_i|U_i) - I(V_{1i};Z_i|U_i)) + 2n\e_n\\
& =  \sum_{i=1}^n(I(V_{1i};Y_i,S_i|U_i) - I(V_{1i};Z_i,S_i|U_i) + I(V_{1i};S_i|Z_i,U_i))+ 2n\e_n\\
& \le \sum_{i=1}^n(I(V_{1i};Y_i,S_i|U_i) - I(V_{1i};Z_i,S_i|U_i) + H(S_i|Z_i,U_i))+ 2n\e_n\\
& \le \sum_{i=1}^n(I(V_{1i};Y_i|U_i, S_i) - I(V_{1i};Z_i,S_i|U_i,S_i) + H(S_i|Z_i,U_i))+ 2n\e_n\\
& \stackrel{(f)}{=}n(I(V_1;Y|U,S) - I(V_1;Z|U,S) + H(S|Z,U))+ 2n\e_n,
\end{align*}
where $(a)$ follows by Fano's inequality; $(b)$ follows from the secrecy condition; $(c)$ and $(d)$ follows the Csisz\'{a}r sum identity; $(e)$ follows from defining $U_i = (Y_{i+1}^n, S_{i+1}^n, Z^{i-1})$ and $V_{1i} = (M,Y_{i+1}^n, S_{i+1}^n, Z^{i-1})$; and $(f)$ follows from setting $Q$ to be a uniform random variable over $[1:n]$, independent of all other variables, and defining $U = (U_Q,Q)$, $V_{1} = (V_{1Q}, Q)$, $S = S_Q$, $Y = Y_Q$ and $Z = Z_Q$. }

For the second upper bound, we have{\allowdisplaybreaks
\begin{align*}
nR &\le I(M;Y^n,S^n) + n\e_n \\
& \stackrel{(a)}{=} I(M;Y^n|S^n) + n\e_n \\
& = \sum_{i=1}^n I(M;Y_i|S^n, Y_{i+1}^n) \\
& \le \sum_{i=1}^n I(M, Y_{i+1}^n, Z^{i-1}, S_{i+1}^n, S^{i-1};Y_i|S_i) \\
& \stackrel{(b)}{=} \sum_{i=1}^n I(V_{2i}; Y_i|S_i) \\
& = nI(V_{2Q};Y|S,Q) \\
& \stackrel{(c)}{\le} nI(V_2;Y|S), 
\end{align*}
where $(a)$ follows from the independence between $M$ and $S^n$; $(b)$ follows from defining \\ $V_{2i} = (M, Y_{i+1}^n, Z^{i-1}, S_{i+1}^n, S^{i-1})$ and $(c)$ follows from defining $V_2 = (V_{2Q}, Q)$.} 

\section{Conclusion} \label{sect:6}
We established bounds on the secrecy capacity of the wiretap channel with state information causally available at the encoder and decoder. We showed that our lower bound  can be strictly larger than the best known lower bound for the noncausal state information case. The upper bound holds when the state information is available noncausally at the encoder and decoder. We showed that the bounds are tight for several classes of wiretap channels. 

We used key generation from state information to improve the message transmission rate. It may be possible to extend this idea to the case when state information is available only at the encoder. This case, however, is not straightforward to analyze since it would be necessary for the encoder to reveal some state information to the decoder (and hence partially to the eavesdropper) in order to agree on a secret key. This may reduce the wiretap coding part of the rate.

\bibliographystyle{IEEEtran}
\bibliography{secrecy}

\appendices 

\section{Appendix: Proof of Proposition~\ref{prop1}} \label{appen1}
1. The proof of this result follows largely from Lemma 2 in Lecture 23 of Lectures on Network Information Theory by El Gamal and Kim~\cite{El-Gamal--Kim2010}. For completeness, we give the proof here. Consider
\begin{align*}
H(K_j|\Cc) &\ge \P\{S^n \in \aep \} H(K_j |\Cc, \Sv(j) \in \aep)\\
&\ge (1-\e'_n) H(K_j|\Cc, \Sv(j) \in \aep).
\end{align*}
Let $P(k_j)$ be the {\em random} pmf
of $K_j$ given $\{\Sv(j) \in \aep\}$, where the randomness is induced by the random bin
assignment (codebook) $\Cc$. 

By symmetry, $P(k_j)$, $k_j \in [1:2^{nR_K}]$, are identically distributed. We express $P(1)$ in terms of a weighted sum of indicator functions as
\[
P(1) = \sum_{s^n \in \aep} \frac{p(s^n)}{\P\{S^n \in \aep \}} \cdot {I}_{\{s^n \in \Bc(1)\}}.
\]
It can be easily shown that
\begin{align*}
\E_{\Cc}(P(1)) &= 2^{-nR_K},\\
\var(P(1)) &= 2^{-nR_K}(1- 2^{-nR_K}) \sum_{x^n \in \aep} \left(\frac{p(s^n)}{\P\{\Sv(j) \in \aep\}}\right)^2 \\
&\le 2^{-nR_K} 2^{n(H(S) + \d(\e))} \frac{2^{-2n(H(S) - \d(\e)) }} {(1- \e'_n)^2} \\
&\le 2^{-n(R_K+H(S) - 4\d(\e))}
\end{align*}
for sufficiently large $n$.  

By the Chebyshev inequality, 
\begin{align*}
\P\{ |P(1) - \E(P(1))| 
\ge \e \E(P(1)) \} &\le \frac{\var(P(1))}{(\e \E(P(1)))^2}\\
&\le \frac{2^{-n(H(S)-R_K - 4 \d(\e)) }}{ \e^2}.
\end{align*}
Note that if $R_K< H(S)-4 \d(\e)$, this probability $\to 0$ as $n \to \infty$. 
{\allowdisplaybreaks
Now, by symmetry 
\begin{align*}
& H(K_1|\Cc, \Sv(j) \in \aep) \\
& = 2^{nR_K} \E(P(1)) \log (1/P(1))) \\
& \ge 2^{nR_K} \P\{|P(1) - \E(P(1))| < \e 2^{-nR_K}  \} 
\E\bigl(P(1)\log (1/P(1)) \,\big|\,
|P(1) - \E(P(1))| < \e 2^{-nR_K} \bigr)\\
& \ge \left(1- \frac{2^{-n(H(S)-R_K - 4\d(\e))}}{\e^2} \right) 
\cdot (nR_K (1-\e)-(1-\e)\log (1+\e)) \\
& \ge n(R_K - \d(\e))
\end{align*}
}for sufficiently large $n$ and $R_K< H(S)-4 \d(\e)$.

Thus, we have shown that if $R_K< H(S)-4 \d(\e)$, $H(K_j|\Cc) \ge n(R_K - \d(\e))$ for $n$ sufficiently large. This completes the proof of part 1 of Proposition~\ref{prop1}. Note now that since $H(S|Z) \le H(S)$, the same results also holds if $R_K \le H(S|Z) -4\d(\e)$.
\medskip

2. We need to show that if $R_K < H(S|Z) - 3\d(\e)$, then $I(K_j; \Zv(j)|\Cc) \le 2n\d(\e)$ for every $j\in [1:b]$.
We have  
\begin{align*}
I(K_j;\Zv(j)|\Cc) &= I(\Sv(j); \Zv(j)|\Cc) - I(\Sv(j);\Zv(j)|K_j, \Cc).
\end{align*}
We analyze the terms separately. For the first term, we have
\begin{align*}
I(\Sv(j); \Zv(j)|\Cc) & = I(\Sv(j),L;\Zv(j)|\Cc) - I(L;\Zv(j)|\Sv(j), \Cc) \\
& \le I(U^n,\Sv(j);Z|\Cc) - H(L|\Sv(j), \Cc) + H(L|\Sv(j), Z^n) \\
& \le nI(U,S;Z) - H(L|\Sv(j), \Cc) + H(L|\Sv(j), Z^n) \\
& \stackrel{(a)}{\le} nI(U,S;Z) - H(L|\Sv(j), \Cc) + n(\Rt - I(U;Z,S) + \d(\e)) \\
& = n\Rt - H(M_{j0}|\Cc) -H(M_{j1}\oplus K_{j-1}|\Cc, M_{j0}) \\
& \quad -H(L|M_{j0}, M_{j1}\oplus K_{j-1}, \Cc) + nI(S;Z) +n\d(\e) \\
& \le n\Rt - nR_0 -H(M_{j1}\oplus K_{j-1}|\Cc, M_{j0}, K_{j-1}) - n(\Rt-R_0-R_K) + nI(S;Z) +n\d(\e) \\
& = nR_K - H(M_{j1}|\Cc, M_{j0}, K_{j-1}) + nI(S;Z) +n\d(\e) \\
& = n(I(S;Z)+\d(\e)), 
\end{align*}
where step $(a)$ follows from application of Lemma 1 which holds since $\Rt-R_0 \ge I(U;Z,S)$.
For the second term we have
\begin{align*}
I(\Sv(j);\Zv(j)|K_j, \Cc) & = H(\Sv(j)|K_j, \Cc) - H(\Sv(j)|\Zv(j), K_j, \Cc) \\
& = H(\Sv(j), K_j|\Cc) -H(K_j|\Cc) - H(\Sv(j)|\Zv(j), K_j, \Cc) \\
& \ge nH(S) -nR_K -H(\Sv(j)|\Zv(j), K_j, \Cc) \\
& \ge n(H(S) - R_K) -H(\Sv(j)|\Zv(j), K_j) \\
& \stackrel{(b)}{\ge} n(H(S) - R_K) - n(H(S|Z) - R_K + \d'(\e)) \\
& = nI(S;Z) -n\d(\e),
\end{align*}
where $(b)$ follows from showing that $H(\Sv(j)|\Zv(j), K_j) \le n(H(S|Z) - R_K + \d(\e))$. This requires the condition $R_K < H(S|Z) - 3\d(\e)$. Combining the bounds for the 2 expressions gives $I(K_j;\Zv(j)|\Cc) \le 2n\d(\e)$.
\medskip

\noindent \textit{Proof of step $(b)$}:
Give an arbitrary ordering to the set of all state sequences $s^n$ with $\Sv(j) = s^n(T)$ for some $T\in [1:2^{n\log|S|}]$. Hence,
$H(\Sv(j)|\Zv(j), K)  = H(T|K,\Zv(j))$.

From the coding scheme, we know that $\P\{(s^n(T), \Zv(j))\in \aep\}\to 1$ as $n \to \infty$. Note here that $T$ is random and corresponds to the realization of $S^n$.

Now, fix $T = t$, $\Zv(j) = z^n$, $K = k$ and define $N(z^n, k, t) := |\lt \in [1: |\aep(S)|]: (s^n(\lt),z^n)\in \aep, \; \lt \neq t, \; s^n(\lt) \in \Bc(k)|$. For $z^n \notin \aep$, $N(z^n, k, t) = 0$. For $z^n\in \aep$, it is easy to show that
\begin{align*}
\frac{|\aep(S|Z)|-1}{2^{nR_K}}\le \E(N(z^n, k, t)) &\le \frac{|\aep(S|Z)|}{2^{nR_K}}, \\
\var(N(z^n, k, t)) &\le \frac{|\aep(S|Z)|}{2^{nR_K}}.
\end{align*}

By the Chebyshev inequality, 
\begin{align*}
\P\{ N(z^n, k,t)\ge  (1+\e) \E(N(z^n, k,t)) \} &\le \frac{\var(N(z^n, k,t))}{(\e \E(N(z^n, k,t)))^2}\\
&\le \frac{2^{-n(H(S|Z) - 3\d(\e)-R_K)}}{ \e^2}.
\end{align*}
Note that $\P\{ N(z^n, k,t)\ge  (1+\e) \E(N(z^n, k,t)) \}\to 0$ as $n \to \infty$ if $R< H(S|Z) - 3\d(\e)$. Now, define the following events
\begin{align*}
\Ec_1 &:= \{(\Sv(j), \Zv(j)) \notin \aep\}, \\
\Ec_2 & := \{N(\Zv(j), K, T)\ge (1+\e)\E(N(\Zv(j), K, T))\}.
\end{align*}

Let $E = 0$ if $\Ec_1^c \cap \Ec_2^c$ occurs and $1$ otherwise. We have
\begin{align*}
\P(E = 1) &\le \P(\Ec_1) + \P(\Ec_2) \\
& \le \P(\Ec_1) + \sum_{(z^n,s^n(t))\in \aep,\; k} p(z^n,t,k)\P\{N(z^n,k,t)\ge (1+\e)\E(N(z^n,k,t))\}\\
& \quad + \P\{(s^n(T),\Zv(j))\notin \aep\}. 
\end{align*}
$\P\{(s^n(T),\Zv(j))\notin \aep\} = \P(\Ec_1)$ and $\P(\Ec_1) \to 0$ as $n\to \infty$ by the coding scheme. For the second term, $\P\{N(z^n,k,t)\ge (1+\e)\E(N(z^n,k,t))\} \to 0$ as $n \to \infty$ if $R< H(S|Z) - 3\d(\e)$. Hence, $\P(E=1) \to 0$ as $n \to \infty$ if if $R< H(S|Z) - 3\d(\e)$.

We can now bound $H(T|K,Z^n)$ by
\begin{align*}
H(T|K,Z^n) &\le 1 + \P(E = 1)H(T|K,Z^n, E= 1) + H(T|K,Z^n, E= 0) \\
&\le n(H(S|Z) -R_K + \d(\e)).
\end{align*}
\medskip

3. To upper bound $I(K_j;\Zv^{j}|\Cc)$, we use an induction argument assuming that $I(K_{j-1};\Zv^{j-1}|\Cc) \le n\d_{j-1}(\e),$ where $\d_{j-1}(\e)\to 0$ as $\e \to 0$. Note that the proof for $j=2$ follows from part 2. Consider{\allowdisplaybreaks
\begin{align*}
I(K_j; \Zv^{j}|\Cc)  & = I(K_j; \Zv(j)|\Cc) + I(K_j; \Zv^{j-1}|\Cc, \Zv(j)) \\
& \stackrel{(a)}{\le} 2n\d(\e) + I(K_j; \Zv^{j-1}|\Cc, \Zv(j)) \\
& = H(\Zv^{j-1}|\Cc, \Zv(j)) -  H(\Zv^{j-1}|\Cc, \Zv(j), K_j) +2n\d(\e)\\
& \le H(\Zv^{j-1}|\Cc) - H(\Zv^{j-1}|\Cc, K_{j-1}, \Zv(j), K_j) +2n\d(\e) \\
& \stackrel{(b)}{=} H(\Zv^{j-1}|\Cc) - H(\Zv^{j-1}|\Cc, K_{j-1}) +2n\d(\e)\\
& = I(K_{j-1}; \Zv^{j-1}|\Cc) + 2n\d(\e) \\
& \stackrel{(c)}{\le} n\d_{j-1}(\e) + 2n\d(\e),
\end{align*}
}where $(a)$ follows from part 2 of the Proposition; $(b)$ follows from the Markov Chain relation $\Zv^{j-1} \to K_{j-1} \to (\Zv(j), K_{j})$; $(c)$ follows from the induction hypothesis. This completes the proof since the last line implies that there exists a $\d'(\e),$ where $\d'(\e) \to 0$ as $\e \to 0,$ that upper bounds $I(K_j; \Zv^{j}|\Cc)$ for $j\in [1:b].$

\section{Proof of Proposition 2}~\label{appen2}
1. We first show that if $R_K < H(S) - 4\d(\e)$, then $H(K_j|\Cc) \ge n(R_K - \d(\e))$. This is done in the same manner as 1 of Proposition 1. The proof is therefore omitted. 
\medskip

2. We need to show that if $R_K < H(S|Z) - 3\d(\e)$, then $I(K_j; \Zv(j)|\Cc) \le 2n\d(\e)$ for every $j\in [1:b]$.
We have  
\begin{align*}
I(K_j;\Zv(j)|\Cc) &= I(\Sv(j); \Zv(j)|\Cc) - I(\Sv(j);\Zv(j)|K_j, \Cc).
\end{align*}
We analyze the terms separately. For the first term, we have
\begin{align*}
I(\Sv(j); \Zv(j)|\Cc) & = I(\Sv(j),L;\Zv(j)|\Cc) - I(L;\Zv(j)|\Sv(j), \Cc) \\
& \le I(U^n,\Sv(j);Z|\Cc) - H(L|\Sv(j), \Cc) + H(L|\Sv(j), Z^n) \\
& \le nI(U,S;Z) - H(L|\Sv(j), \Cc) + H(L|\Sv(j), Z^n) \\
& \stackrel{(a)}{\le} nI(U,S;Z) - H(L|\Cc) + n(\Rt - I(U;Z,S) + \d(\e)) \\
& = n\Rt - H(K_{(j-1)d}|\Cc) - H(K_{(j-1)m}\oplus M_j|\Cc) + nI(S;Z) \\
& \quad - H(L|K_{(j-1)m}\oplus M_j, K_{(j-1)d}) + n\d(\e) \\
& \stackrel{(b)}{\le} n(\Rt -R_d  - R - \Rt + R_d +R + 2\d(\e))+nI(S;Z)\\
& = n(I(S;Z)+2\d(\e)), 
\end{align*}
where step $(a)$ follows from application of Lemma 1, which holds from the condition that $\Rt \ge I(U;Z,S)$, and the fact that $\Sv(j)$ is independent of $L$. Step $(b)$ follows from part 1 of Proposition~\ref{prop2}: $H(K_{j-1}|\Cc) \ge n(R_K - \d(\e))$, which implies that $H(K_{(j-1)d}|\Cc) \ge n(R_d - \d(\e))$. Note we implicitly assumed $j\ge 2$. The case of $j=1$ is straightforward, since $H(L|\Cc) = n\Rt$ by the fact that we transmit a codeword picked uniformly at random. 

The proof that $I(\Sv(j);\Zv(j)|K_j, \Cc)\ge nI(S;Z) - n\d(\e)$ follows the same steps as the proof of part 2 of Proposition~\ref{prop1} and requires the same condition that $R_K < H(S|Z) - 3\d(\e).$ 
\medskip

3. Part 3 of the Proposition is proved in the same manner as part 3 of Proposition~\ref{prop1}. 

\section{Proof of Proposition 3}~\label{appen3}
1. We first show that if $R_K < H(S) - 4\d(\e)$, then $H(K_j|\Cc) \ge n(R_K - \d(\e))$. This is done in the same manner as 1 of Proposition 1. The proof is therefore omitted. 
\medskip

2. We need to show that if $R_K < H(S|Z,V) - 3\d(\e)$, then $I(K_j; \Zv(j)|\Cc) \le n\d(\e)$ for every $j\in [1:b]$.
We have  
\begin{align*}
I(K_j;\Zv(j)|\Cc) &\le I(K_j;\Zv(j), U^n|\Cc) \\ 
& =I(\Sv(j); \Zv(j), U^n|\Cc) - I(\Sv(j);\Zv(j),U^n|K_j, \Cc).
\end{align*}
We analyze the terms separately. For the first term, we have
\begin{align*}
I(\Sv(j); \Zv(j), V^n|\Cc) &= I(\Sv(j); \Zv(j)|V^n, \Cc) \\
& = \sum_{i=1}^n (H(\Zv_i(j)|\Cc, V^n, \Zv^{i-1}(j)) - H(\Zv_i(j)|\Cc, V^n, \Sv(j), \Zv^{i-1}(j))) \\
& \le \sum_{i=1}^{n}(H(\Zv_i(j)|\Cc, V_i) - H(\Zv_i(j)|\Cc, V_i, \Sv_i(j))) \\
& \le n(H(Z|V) - H(Z|V,S)) \\
& = nI(Z;S|V) = nI(Z,V;S).
\end{align*}

For the second term, we have
\begin{align*}
I(\Sv(j);\Zv(j), V^n|K_j, \Cc) & = H(\Sv(j)|K_j, \Cc) - H(\Sv(j)|\Zv(j), V^n, K_j, \Cc) \\
& = H(\Sv(j), K_j|\Cc) -H(K_j|\Cc) - H(\Sv(j)|\Zv(j), V^n, K_j, \Cc) \\
& \ge nH(S) -nR_K -H(\Sv(j)|\Zv(j), V^n, K_j, \Cc) \\
& \ge n(H(S) - R_K) -H(\Sv(j)|\Zv(j),V^n, K_j) \\
& \stackrel{(b)}{\ge} n(H(S) - R_K) - n(H(S|Z,V) - R_K + \d'(\e)) \\
& = nI(S;Z,V) -n\d(\e),
\end{align*}

The proof of step $(b)$ follows the same steps as in the proof of part 2 of Proposition~\ref{prop1}. We can show that step $(b)$ holds if $R_K < H(S|Z,V) - 3\d(\e).$

Combining the two terms then give the required upper bound which completes the proof of Part 2.
\medskip

3. Part 3 of the Proposition is proved in the same manner as part 3 of Proposition~\ref{prop1}.

\end{document}